\documentclass{article}

\usepackage{PRIMEarxiv}

\usepackage[utf8]{inputenc} 
\usepackage[T1]{fontenc}    
\usepackage{hyperref}       
\usepackage{url}            
\usepackage{booktabs}       
\usepackage{amsfonts}       
\usepackage{nicefrac}       
\usepackage{microtype}      
\usepackage{lipsum}
\usepackage{fancyhdr}       
\usepackage{graphicx}       
\graphicspath{{media/}}     
\usepackage{pgfplots}
\usepackage{atbegshi}
\usepackage{fancyhdr}
\usepackage{caption}
\usepackage[multiple]{footmisc}
\usepackage{subfig}
\usepackage{graphicx}
\usepackage{amsmath}
\usepackage{multirow}
\usepackage{amssymb}
\usepackage{subcaption}
\usepackage{booktabs}
\usepackage{xcolor}
\usepackage[normalem]{ulem}
\title{Social-Sensor Identity Cloning Detection Using Weakly Supervised Deep Forest and Cryptographic Authentication
}

\author{
Ahmed Alharbi\\
 College of Computer Science and Engineering,\\
 Taibah University, Medina, Saudi Arabia\\    
\texttt{email:atharbi@taibahu.edu.sa} \\
  \AND
 Hai Dong, Xun Yi \\
 School of Computing Technologies\\
  RMIT University, Melbourne, Australia\\
 \texttt{email:\{hai.dong,xun.yi\}@rmit.edu.au} \\
}

\begin{document}
\maketitle

\begin{abstract}
Recent years have witnessed a rising trend in social-sensor cloud identity cloning incidents. However, existing approaches suffer from unsatisfactory performance, a lack of solutions for detecting duplicated accounts, and a lack of large-scale evaluations on real-world datasets. We introduce a novel method for detecting identity cloning in social-sensor cloud service providers. Our proposed technique consists of two primary components: 1) a similar identity detection method and 2) a cryptography-based authentication protocol. Initially, we developed a weakly supervised deep forest model to identify similar identities using non-privacy-sensitive user profile features provided by the service. Subsequently, we designed a cryptography-based authentication protocol to verify whether similar identities were generated by the same provider. Our extensive experiments on a large real-world dataset demonstrate the feasibility and superior performance of our technique compared to current state-of-the-art identity clone detection methods.
\end{abstract}

\keywords{Identity cloning detection \and Social-sensor cloud services \and  Non-privacy-sensitive user profile features \and   Weakly supervised deep forest model \and  Cryptography-based authentication}

\section{Introduction}\label{sec:introduction}
Social sensing is a paradigm that allows multiple \textit{social sensors} (e.g., humans with smart devices) to collect data \cite{rosi2011social}. This sensed data is called \textit{social-sensor data} that is in a variety of forms (e.g., texts, images, etc.) and stored in \textit{social-sensor clouds} (i.e., social media platforms) \cite{aamir2017social,aamir2017social1}. Examples of social-sensor data include  Twitter posts and Facebook status. Social-sensor clouds can be viewed as an essential open channel for social-sensor cloud users to share their personal experiences about special events, such as accidents and public events \cite{rosi2011social}, \cite{aamir2017social}. 
Social-sensors can publish thousands or even millions of posts across social-sensor clouds. This type of data can be abstracted as \textit{social-sensor cloud services} (SocSen services) with \textit{functional} (e.g., location, time, etc.) and \textit{non-functional} aspects (e.g., trust, resolution, etc.) \cite{aamir2017social,aamir2017social1}. It can portray unique events from a variety of perspectives, including where, when, and what \cite{9001217}. 

The prosperity of social-sensor clouds has attracted the attention of cybercriminals in recent years. These cybercriminals frequently attempt to exploit the identities of \textit{SocSen service providers}
 (i.e., users who publish social-sensor data from social media platforms) 
to deceive consumers in a number of various ways \cite{gupta2013faking,mendoza2010twitter}. 
Identity cloning is a way to attack the identities of SocSen service providers, in which an attacker creates a fake profile on a social-sensor cloud based on the identity information of a provider. 
Identity cloning is in two major forms: \textit{single-site identity cloning} and \textit{cross-site identity cloning} \cite{bilge2009all}. The former applies to a scenario in which the identity of a SocSen service provider is cloned and registered in the same social-sensor cloud by an intruder, while the latter refers to a scenario where the identity of a SocSen service provider is copied from a cloud to another. Our project focuses on the detection of single-site identity cloning. In recent years, we have noticed a number of famous cases related to single-site identity cloning. For example, Facebook CEO Mark Zuckerberg's Facebook account was cloned and utilized in a financial fraud\footnote{https://www.nytimes.com/2018/04/25/technology/fake-mark-zuckerberg-facebook.html}. Additionally, Russian President Vladimir Putin's cloned Twitter account had attracted over 1 million followers\footnote{https://www.abc.net.au/news/2018-11-29/twitter-suspends-account-impersonating-vladimir-putin/10569064}.

Most social-sensor clouds do not have mechanisms for automated identity cloning detection. For instance, platforms like Instagram and Twitter only conduct identity cloning investigations after receiving complaints from end-users\footnote{https://help.twitter.com/en/rules-and-policies/twitter-impersonation-policy}\footnote{https://help.instagram.com/446663175382270}. Additionally, many existing identity cloning detection techniques depend on a mix of privacy-sensitive (e.g., full name, date of birth) and non-privacy-sensitive user profile features (e.g., screen name, profile descriptions) \cite{devmane2014detection,kamhoua2017preventing,kontaxis2011detecting}. However, these methods are inapplicable to most third-party websites and applications for detecting cloned identities. This is because most third-party platforms that rely on social media APIs for user authentication (i.e., social login) are restricted from accessing privacy-sensitive user data. On the other hand, non-privacy-sensitive user profile information is generally accessible through APIs to third-party platforms. Therefore, there is a strong need to explore new techniques for identity cloning detection that rely exclusively on non-privacy-sensitive user profile features.

Further, most existing identity cloning detection techniques 
\cite{devmane2014detection,goga2015doppelganger,jin2011towards} rely on relatively simple  metrics, such as Jaro–Winkler distance and Term Frequency — Inverse Document Frequency (TF-IDF) based cosine similarity, for inter-identity feature similarity measure.
These metrics rely on word frequency or character distance and \textit{cannot fully capture the semantics of  words and strings}. For instance, the (gender-based) similarity of \textit{king} and \textit{man} cannot be measured using the previous metrics. Deep learning (DL) techniques, such as Bidirectional Encoder Representations from Transformers (BERT), provide embeddings that enable us to create more than one representation for a word based on its context of usage \cite{devlin2018bert}. In this regard, DL can capture the semantics of words and strings.
In addition, existing identity cloning detection techniques employ conventional machine learning \cite{goga2015doppelganger}. Conventional machine learning uses shallow architectures and is incapable of handling large-scale data sources. In this context, DL has been acknowledged as a powerful tool for large data processing and analytics in a variety of domains \cite{najafabadi2015deep, 9599374, 8653299, 8523670, 9778269, 9590330, 4635214}. DL architecture can transform and represent information at multiple stages.

Against the issues above, we previously proposed 1) an unsupervised identity cloning detection approach based on non-privacy-sensitive user profile features \cite{alharbi1234} and 2) a deep learning model for identity cloning detection \cite{alharbi2021privacy, 10715673}. However, our previous works and the existing approaches still suffer from the following major limitations:
\begin{itemize}
    \item \textit{Unsatisfactory performance.} The performance of the existing identity cloning detection techniques still has sufficient space for improvement.
    \item \textit{Lack of solutions for duplicated accounts.} Most of the existing works only focus on detecting similar identities and assume that all of them are cloned. However, it is quite common that many providers create more than one similar identity in a social-sensor cloud using different emails. This kind of identities is also termed as \textbf{duplicate accounts}. These identities are obviously not cloned/fake ones created by attackers. 
    
    \item \textit{Lack of large-scale evaluations on real-world datasets}. The existing techniques were only tested in simulated or small-scale datasets.
\end{itemize}

To address the aforementioned limitations, we propose a novel technique for SocSen service providers' identity cloning detection based on non-privacy-sensitive user features. Our proposed technique comprises two main components, namely, 1) a similar identity detection approach to detect a pair of accounts that might contain a cloned account and its victim and 2) a cryptography-based cloned identity authentication protocol to validate whether  a pair of identified accounts were created by an identical provider. Our main contributions are summarised below.
\begin{itemize}
      \item[--] We designed a weakly supervised deep forest model to further improve the similar identity detection performance.
      First, we trained a deep forest classifier to predict the weak labels. Then we used the generated weak labels to train the deep forest classifier.
      \item[--] {We designed a cryptography-based authentication protocol to validate whether or not a pair of identified similar accounts contain a cloned account. This protocol requests the newer account from an account pair to decrypt two random messages respectively encrypted by the newer and older accounts' public keys for authentication. This is expected to identify the most common noises during cloned account detection, which is that a SocSen service provider creates duplicate accounts in a social-sensor cloud}.
    \item[--] We conduct a more comprehensive evaluation on a large real-world dataset. The dataset contains more than 500K accounts collected from Twitter. The experimental results show that our proposed technique outperforms the state-of-the-art identity cloning detection techniques on precision and F1-score.
\end{itemize}
The remainder of the paper is structured as follows. Section \ref{re_work} reviews the current literature in SocSen services, identity cloning detection and identity-based encryption. Section \ref{over_sol} presents the details of  the proposed framework. Section \ref{eval} explains the experimental design and result analysis of our proposed framework. Section \ref{concl} concludes the paper.
\section{Related Work} \label{re_work}
This section analyzes and reviews the current literature in SocSen services, identity cloning detection and identity-based encryption.
\subsection{Social-Sensor Cloud Services}

SocSen services provide simple access to the management of social-sensor cloud data for the development of scene analysis applications. Aamir et al. \cite{aamir2017social,aamir2017social1} introduced a collection of SocSen service selection frameworks for user to query scene-related social media images, where SocSen services abstract the functional and non-functional properties of social media images. These frameworks aim to organize social media images spatially, temporally, and contextually. Aamir et al. \cite{9001217,aamir2018social} further proposed a set of SocSen scene analysis service composition models. These models reconsruct tapestry scenes by using the spatio-temporal, textual and image features of the SocSen services.

\subsection{Identity Cloning Detection Techniques}
A variety of approaches have been proposed for fake user accounts and spammers detection on social media platforms according to a survey  \cite{alharbi2021social}. These approaches mostly utilise behavioural features of users (e.g. writing styles, friendship lists, etc) to identify whether or not a user account is trustworthy \cite{masood2019spammer,zheng2015detecting}. However, the behavioural profile features cannot be effectively employed for identity cloning detection, since an adversary can easily mimic the behavioural profile features. This requires us to explore new features that can adequately describe an account pair. Other researchers, on the other hand, employ social network trust links between users.  These approaches assume that a spammer/fake user cannot develop an arbitrary number of trusted links with legitimate users \cite{al2017sybil,masood2019spammer}. However, in the context of identity cloning, this assumption may not hold true, as adversaries can attempt to clone legitimate users profiles. Thus, cloned accounts can more easily establish trust with legitimate users.

Several approaches have been proposed for detecting social media identity cloning. Vyawahare and Govilkar \cite{vyawahare2022fake} developed a method to detect fake and cloned profiles by extracting key attributes (e.g., username, friend count, gender) and calculating a similarity index. Profiles with high similarity scores above a threshold are flagged as potential clones. Jethava and Rao \cite{jethava2022novel} introduced a defensive approach to protect against identity cloning. Their method uses similarity measures (e.g., attribute and friend list) to differentiate between cloned and legitimate users. The approach is implemented on the social app server, where friendship requests are checked for authenticity before being approved. Kontaxis et al. \cite{kontaxis2011detecting} introduced a mechanism by which users can ascertain if they have fallen in a cloned identity attack.  Jin et al. \cite{jin2011towards} studied the behaviour of attackers for identity cloning. They presented two approaches to identify suspicious profiles based on profile similarity. Goga et al. \cite{goga2015doppelganger} proposed an impersonation detection technique. The proposed technique determines whether a pair of accounts are duplicate  accounts. It uses a Support Vector Machine classifier to detect impersonation attacks. Kamhoua et al. \cite{kamhoua2017preventing} compared user profiles across social media platforms to prevent cloned identities. A hybrid string-matching similarity technique is employed to determine the extent of profile similarity. Devmane and Rana \cite{devmane2014detection} developed a technique to detect cloned identities in both single- and cross-platforms. They developed a technique to search for similar user accounts.

Most of the current works \cite{devmane2014detection,kamhoua2017preventing,kontaxis2011detecting} detect cloned accounts by analysing both privacy-sensitive and non-privacy-sensitive user profile features. Many third-party websites use social media profiles to verify users. These third-party websites are unable to gain access to privacy-sensitive user profile features through social media APIs. Therefore, those techniques cannot be applied by those third-party websites. In addition, all the aforementioned recent works have not considered solutions to distinguish between the cloned and self-duplicated accounts. This inspires us to investigate identity-based encryption for self-duplicated account authentication.

\subsection{Identity-Based Cryptography}

Identity-based cryptography is a type of asymmetric cryptography. It permits the generation of a public key from an arbitrary string (e.g. user's identity). Unlike asymmetric cryptography, it requires the generation of both public and private keys concurrently \cite{shamir1984identity}. Identity-based cryptography simplifies the process of key management. It requires a directory for storing authenticated public system parameters of a Central Authority (CA), which is significantly less cumbersome than keeping a directory for storing all users' public keys \cite{gorantla2005survey}. However, identity-based cryptography requires a secure channel for key distribution between the CA and users. Therefore, in our proposed cryptography-based authentication approach, we encrypt the generated private key for a user using a shared session key before the CA distribute the private key.



A variety of identity-based encryption (IBE) schemes have been proposed. Boneh and Franklin \cite{boneh2001identity} proposed an IBE scheme that is both practical and secure. The proposed scheme is indistinguishably safe against attacks that use adaptively selected ciphertexts (i.e. IND-ID-CCA secure). Cocks \cite{cocks2001identity} proposed an IBE scheme using quadratic residues. However, the proposed solution lacks formal security proofs and is very inefficient in terms of bandwidth needs. Sahai and Waters \cite{sahai2005fuzzy} proposed a fuzzy IBE scheme. The proposed scheme permits the encryption of data utilising biometric input as a public key. We utilise the Boneh-Franklin IBE scheme in our proposed cryptography-based authentication approach. The Boneh-Franklin IBE scheme is provably secure and efficient \cite{gorantla2005survey}. The result is based on novel assumptions about the difficulty of issues in specific elliptic curve groups.


\section{Solution}\label{over_sol}
This section presents a comprehensive overview of the proposed technique and its key components.
\subsection{Solution Overview} \label{sol_over}
\begin{figure*}[]
  \centering
  \includegraphics[width=\textwidth]{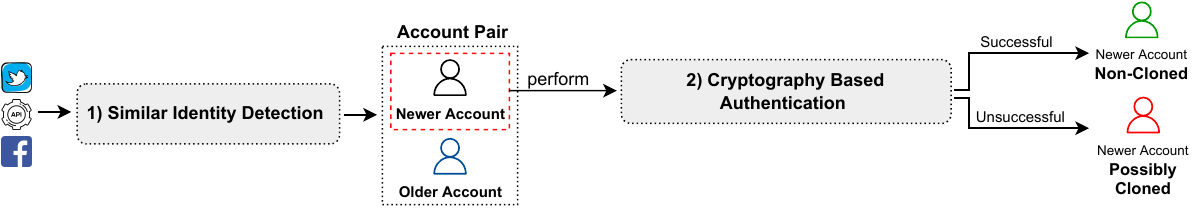}
  \caption{The overview of the proposed solution.}
  \label{overview}
\end{figure*}

We propose a SocSen service provider identity cloning detection  technique. The proposed approach comprises two main components, namely,  1)  a similar identity detection approach and 2) a cryptography-based authentication protocol. Figure \ref{overview} shows the overview of the proposed technique. Given a set of social media (e.g., Twitter, Instagram, etc.) accounts\footnote{We use the terms \textit{SocSen service providers} and \textit{social media users} interchangeably throughout the paper unless specified otherwise.}, we aim to detect  if there are possible identity cloning attacks. Firstly, we employ the similar identity detection approach to detect similar account pairs. The proposed similar identity detection is based on weakly-supervised learning. Once an account pair is detected, we perform a cryptography-based authentication process to the newer account in the account pair. The primary goal of this component is to investigate if the account pair is created by 
the same user. The cryptography-based authentication generates a binary result. If the authentication is successful, it validates that the newer account is not cloned; otherwise the account pair is determined to be highly likely cloned.
\subsection{Similar Identity Detection}\label{sim}
We introduce an overview of our proposed similar identity detection approach and its key components. 
\subsubsection{System Design}
\begin{figure*}[]
  \centering
  \includegraphics[width=\textwidth]{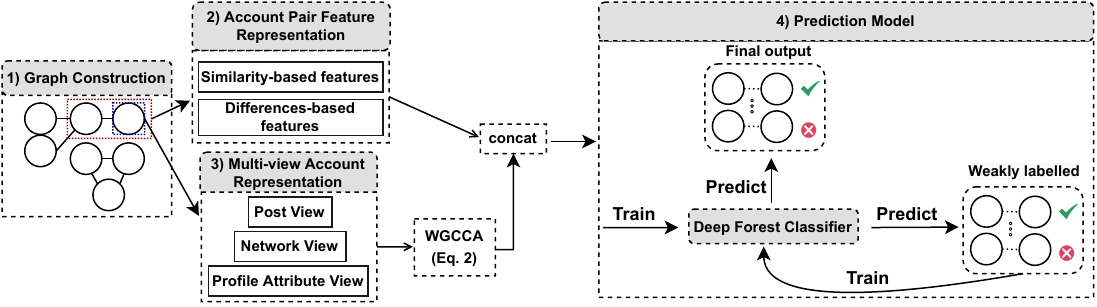}
  \caption{The workflow of the similar identity detection technique}
  \label{over}
\end{figure*}

The proposed similar identity detection approach is illustrated in Figure \ref{over}. It comprises four main components: 1) graph construction (GC); 2) an account pair feature representation; 3) a multi-view account representation and 4) a prediction model. First, the GC seeks to construct an undirected graph from a given set of social media users to locate similar accounts pairs. Second, we extract two categories of non-privacy-sensitive user features between the pair of accounts. Third, we create a multi-view account representation from multiple non-privacy-sensitive views for each account in the account pair.  The account pair feature representation and the multi-view account representation are then concatenated. Next, we aim to verdict whether the pair of accounts possibly consist of a cloned account and a victim account based on a prediction model. In this regard, we first train a DF model to generate weak labels. Once we generate the weak labels, we train another  DF model using the weak labels. The DF model will eventually make the prediction. The following subsections depict these four components in detail.
\subsubsection{Graph Construction}
Our goal is to build an undirected graph connecting among social media accounts, in which each pair of mutually connected nodes indicates a victim account and a possibility cloned account. The username or screen name of a cloned account is highly likely to be same as or similar to the victim account \cite{goga2015doppelganger}. As a result, this graph links nodes according to the similarity of the username and screen name. We connect two nodes through an edge only if the similarity score of their username or screen name is  higher than a predefined threshold $\delta$. This graph can generate all potential account pairs (a victim account and its clone) in the dataset with a proper threshold ($\delta$) value. We elaborate the procedure of choosing a suitable $\delta$ value in the upcoming experiments.

\subsubsection{Account Pair Feature Representation}
After constructing the undirected graph, we extract a group of non-privacy-sensitive user features for each  account pair. All these features can directly be obtained or calculated based on the information extracted from the APIs of mainstream social media platforms, such as Twitter and Facebook. We categorize these features into similarity-based and difference-based features. 
\\(1) \textbf{Similarity-based features}\\ The similarity-based features compare the textual similarity of certain non-privacy-sensitive features (e.g., usernames, screen names, descriptions and locations) between two accounts in each pair. 
We describe the procedure for computing the similarity-based features as follows.

\textbf{Screen name, username and location similarity: }The Jaro–Winkler string similarity (JS) is one of most effective name similarity metric 
\cite{christen2012data,cohen2003comparison}. Therefore, we use JS to measure the textual similarity between two screen names, usernames or locations.

\textbf{Description similarity:} A user account often contains a brief textual description of the user's affiliations, interests, occupations, etc.  We compute the description similarity between a pair of accounts.  First, we pre-process the description text by removing stop words and punctuation and converting it to lowercase.  We then convert the text into vectors by using the term frequency-inverse document frequency (TF-IDF) metric. Next, we calculate the similarity score between the pair of accounts using cosine similarity. 
\\(2) \textbf{Differences-based features}\\
The differences-based features are utilised to compare the general profile features (e.g., posts counts, friends counts, etc.)  of each account pair. We assume that an account pair that possibly contains a cloned account and its victim has a high difference in these general profile features based on the research outcomes of \cite{goga2015doppelganger}. 
For instance, a high difference score of the tweet counts between two similar accounts can possibly indicate a victim account and its replica.
Altogether, we adopt 10 features across the two aforementioned categories. Table \ref{feature} summarizes all the features used for an account pair.
\begin{table*}[]
\caption{Account pair features and descriptions}
\label{feature}
\resizebox{\textwidth}{!}{%
\begin{tabular}{p{0.22\textwidth}|l|p{0.18\textwidth}|p{0.43\textwidth}}
\toprule
\textbf{Feature category}& No. & \textbf{Features} & \textbf{Description} \\ \toprule
 \multirow{2}{*}{Similarity-based features}&1 & Screen name similarity  & The similarity score of screen name between two accounts.  \\  
                   &2& Location similarity &  The similarity score of locations  between two accounts.   \\ 
                  &3 & Username similarity  & The similarity score of usernames between two accounts. \\  
                  &4& Description similarity & The similarity score of descriptions between two accounts.  \\ 
                  &5& Followers Ratio  & The ratio of the number of followers between two accounts. \\ \toprule

\multirow{2}{*}{Differences-based features}&6 &  Followers differences &  The account difference in  the number of followers.  \\ 
                  &7& Account age differences  &  The account difference in account age. \\
                  &8& Tweets differences   &  The account difference in the number of tweets.  \\ 
                  &9& Favorite differences  & The account difference in the number of favorite counts.   \\
                  &10& Friends differences  & The account difference in the number of friends.   \\
                    \toprule
\end{tabular}%
}
\end{table*}
\subsubsection{Multi-view Account Representation}
We aim to create a multi-view account representation for each account in an account pair. The multi-view account representation combines a variety of   views that resemble the non-privacy-sensitive profile features of the account. Here we model three views for a user account: 1) post, 2) network and 3) profile attribute views. These views can accurately represent an account, which are likely mimicked by adversaries. 
Next, we employ weighted generalized canonical correlation analysis (wGCCA) to learn a single embedding from the aforementioned views. This method is effective in capturing the shared structure across multiple views while maximizing the correlation between them \cite{hotelling1992relations}.
The  views are explained below.
\\(1) \textbf{Post View}\\
For each account in the undirected graph, we extract the pre-trained language representation of the account's posts. We obtain the vector-space representations of user posts using Sentence-BERT (SBERT) \cite{reimers-2019-sentence-bert}, which is a variation of the bidirectional encoder representations from transformers (BERT) \cite{devlin2018bert}. 
We gathered $n$ publicly available posts for each account $u$, represented as $T = (t_1, ..., t_n)$. Each post $t_{i} $($i \in 1,..,n$) is represented by the SBERT representation. First,  post $t_i$ is tokenized into a list of words $w_i$. Then special markers [CLS] and [SEP] are added to mark the beginning and ending of a sentence respectively. The BERT layer uses a set of tokenized words to embed fixed-length sentences. The pooling layer then generates $t$ representations using mean aggregation, which outperforms max and CLS aggregation \cite{reimers-2019-sentence-bert}. The output dimension of each post is 385, which is BERT's default output size. Finally, we compute the mean of all posts' representation $T$ for the user account.
\\(2) \textbf{Network View}\\
A network of user accounts is a cluster of individuals who engage in a social graph. Social media users can engage in various ways, such as befriending, posting, etc. Here we employ two kinds of engaging networks: friend and follower networks. The friend network refers to the accounts (e.g., a superstar or friend) followed by an account and thus these accounts' posts appear on the account's feed. 
The follower network refers to the accounts that follow another account and those accounts can see this account's posts on their feed. 
We use Node2Vec  to learn the network representation. Node2Vec is well recognized as an efficient technique for network representation \cite{10.1145/2939672.2939754}. It uses a biased random walk to maximise the log probability of accounts that share an edge.
\\(3) \textbf{Profile Attribute View}\\
We use 12 public profile  attributes to generate the final view of each account. These public profile  attributes are all non-privacy-sensitive user features that can portray the behaviours and activities of an account. For instance, the post count may reflect the activity of the account, while the follower count may indicate the reputation of the account. The 12 public profile  attributes are presented in Table \ref{attributes}.
\begin{table*}[]
\caption{The used non-privacy-sensitive user features for the profile attributes view and their descriptions. }
\label{attributes}
\resizebox{\textwidth}{!}{%
\begin{tabular}{l|p{0.18\textwidth}|p{0.79\textwidth}}
\toprule
 \textbf{No.} & \textbf{Features} & \textbf{Description}  \\ \toprule
 
1 & Follower count & The amount of the user account's followers.  \\ 
2 & Favorite count & The amount  of the user account's liked tweets.  \\ 
3 & Tweet count & The amount  of user account's posts/tweets.  \\ 
4 & Friend count & The amount  of the user account's friends.  \\
5 & List count & The amount  of public lists of which the account is a member.  \\ 
6 &  Account age & The age of user account since registration.  \\ 
7 & Profile background & A binary value indicating if the user account's backdrop or theme has been updated.  \\ 
8 &  Screen name length & The user account's screen name length.  \\ 
9 & Profile image & A binary value indicating whether the user account has a uploaded profile image or a default image.  \\ 
10 &  Description length & The user account's description  length.  \\
11 &  Has profile description & A binary value indicating whether or not the user account's profile includes a description.  \\ 
12 &  Profile URL & A binary value indicating whether or not the user account's profile includes a URL.  \\ 
 \toprule
\end{tabular}
}
\end{table*}
\textbf{\\Embedding Learning Model}\\
The previously mentioned views may provide comprehensive knowledge that can assist in detecting cloned accounts. Employing each view individually might result in knowledge gaps. One strategy is to concatenate all the previously mentioned views. Nevertheless, the view concatenation may generate a large account representation and cause overfitting with limited training datasets. 
It may also omit valuable information, as these views might have unique statistical features \cite{xu2013survey}. For example, each view might have different data patterns (e.g., skewness, uniform, etc.). Accordingly, we employ an alternative approach called generalised canonical correlation analysis (GCCA). GCCA learns a single embedding from a set of views. GCCA has many variants.
Here we employ Carroll’s GCCA, since it is based on a computationally simple and efficient eigenequation \cite{carroll1968generalization}. The objective of GCCA can be formulated as follows.
\begin{equation}
\label{gcca}
    arg \min_{G_i,U_i} \sum_i \parallel G-X_iU_i \parallel_F^2 \qquad s.t. G' G = I 
\end{equation}
where \(X_i \in \mathbb{R}^{\ n\times d_i}\) is the $i^{th}$ view's data matrix, $G \in \mathbb{R}^{\ n\times k}$ includes all learned account embedding and $U_i \in \mathbb{R}^{\ d_i\times k}$ maps the latent space with the observed view $i$. Each view, however, may contain varying degrees of knowledge required to detect cloned accounts. Thus, we use weighted GCCA (wGCCA) that adds weight $w_i$ to each of the views $i$ in Equation \ref{gcca}, as shown in Equation \ref{wgcca}:
\begin{equation}
\label{wgcca}
    arg \min_{G_i,U_i} w_i\sum_i \parallel G-X_iU_i \parallel_F^2 \; s.t. G' G = I, w_i\geq0 
\end{equation}
where $w_i$ is the weight of a view,  which indicates its importance. The columns of $G$ are the eigenvectors of $\sum_i w_i X_i({X_i}'X_i)^{-1}{X_i}'$. The solution is $U_i=({X_i}'X_i)^{-1}{X_i}'G$.

\subsubsection{Prediction Model}
We aim to predict if a pair of similar accounts possibly comprise a cloned account and its victim. We first train a Deep Forest (DF) model  to generate weak labels. Once the weak labels are generated,  we employ them to train a new DF model to make the prediction.
\\(1) \textbf{Deep Forest Model}\\\label{df}
We first concatenate the account pair feature representation and the multi-view account representation to generate the final accounts pair representation (\(A_i\)). 
\begin{equation}
\label{concat}
    y = classifer(concat(F, wgcca) 
\end{equation}
where \(F\) is a feature vector \(F\) ($\in \mathbb{R}^{10}$), in which each feature $F_i$ in $F$ represents an augmented feature obtained from the account pair feature representation and \(wgcca\) is the multi-view account representation embedding.

We utilise the DF model to predict if an account pair possibly includes a cloned account and its target account. The DF model is an ensemble decision tree framework that works effectively with less data and fewer hyperparameters. The structure of the DF model follows a cascaded pattern, in which the input in each layer is a concatenation of the feature vectors produced by the previous layer
\cite{zhou2017deep}.

\begin{figure*}[]
  \centering
  \includegraphics[width=\textwidth]{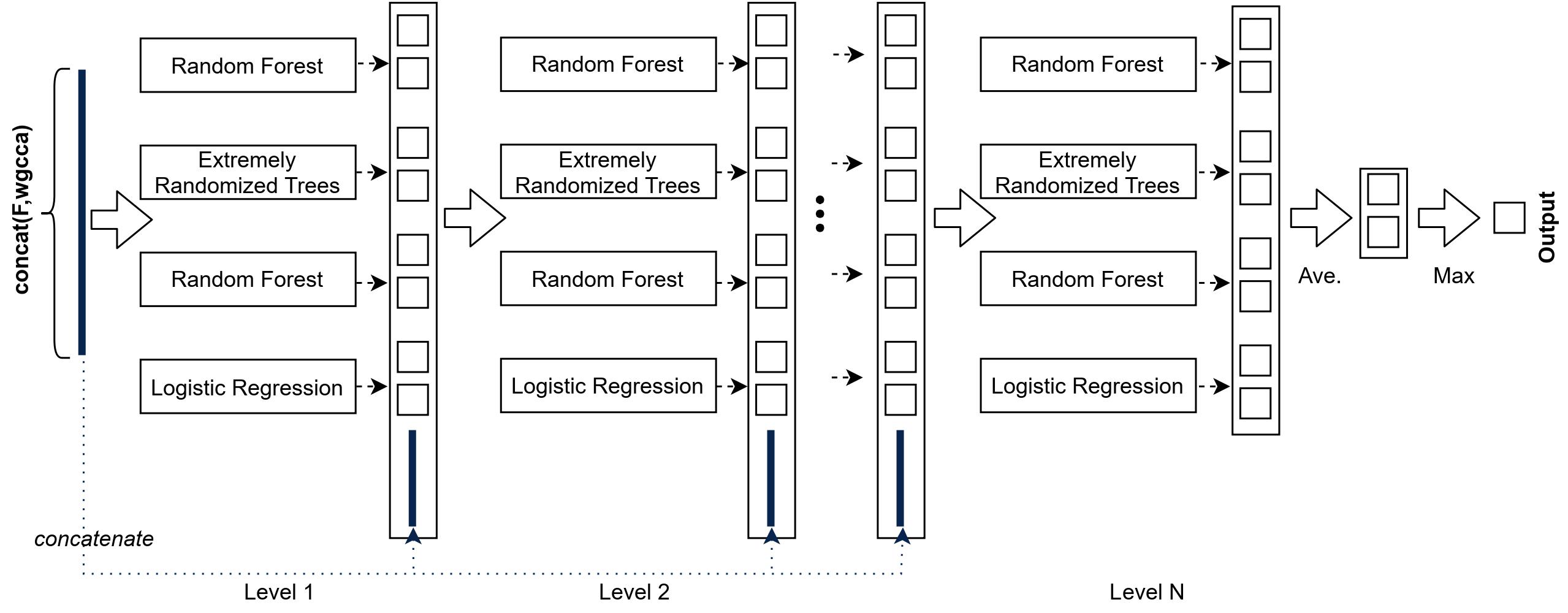}
  \caption{The proposed DF model architecture.}
  \label{deep_forest}
\end{figure*}

Diversity is strongly advised for constructing an ensemble model \cite{zhou2012ensemble}. Hence, we employ two random forests (RF), extremely randomized trees (ERT) and logistic regression (LR) in the proposed DF model. Figure \ref{deep_forest} depicts the proposed DF architecture, which contains multiple  self-adaptive layers. We elaborate on the procedure of choosing the proposed DF architecture in  Sec. \ref{RQ2_result}. The RF model is an ensemble classifier that integrates a large number of decision trees and averages their results to enhance the prediction performance \cite{breiman2001random}. The ERT model is similar to the RF model. However, they differ in how splits are determined. A RF can be divided in terms of trees, while the ERT splits randomly \cite{geurts2006extremely}. 

The model is fed with the concatenated feature vector $concat(F, wgcca)$ . 
Herein, we aim to obtain a binary value $y$ indicating whether or not an account pair highly likely contains a cloned account and its victim. In this regard, each of the employed models (i.e., 2 RFs, ERT, LR) generates a binary value separately. Thus, the   input for the next layer is a set of $8$ ($= 2\times4$)
augmented features. The class vector is generated using $k$-fold cross-validation $(k=5)$. This generates $k-1$ class vectors, which are then averaged to generate the final class vector as the augmented features for the cascade's subsequent layer. After adding one extra layer, the validation set is used to measure the prediction performance of the whole cascade. If there is no substantial improvement in class prediction performance, the cascade growth is immediately terminated. As a result, it can be claimed that the number of cascaded layers is self-determined \cite{zhou2017deep}.
\\(2) \textbf{Weak Label Generation}\\
Weakly supervised learning is an emerging research area. It aims to leverage limited, noisy, or imprecise inputs to produce labels for large-scale training data in a supervised learning setting. Weakly supervised learning can take three different shapes: incomplete, inexact and inaccurate supervision \cite{zhou2018brief}. Incomplete supervision refers to the situation in which only a fraction of training data is labelled, such as active learning and semi-supervised learning. Inexact supervision arises from the fact that the training data is provided with only coarse-grained labels. 

Inaccurate supervision (i.e. weak labels) refers to a scenario in which the training data is not always accurate. In other words, some labelled data may contain errors. Weak labels can be utilised with the notion that they are imprecise but may be used to build a robust prediction model \cite{zhou2018brief}.
Weak labels may reduce the cost of building ground truth datasets that are prohibitively expensive or impractical. 
{It is common that large-scale identity cloning datasets often contain noisy samples (e.g., duplicated accounts). In addition, the ground truth is usually difficult to be obtained for such large-scale datasets since social media providers do not usually disclose it. Consequently, creating datasets of weak labels may benefit the training of DL models that usually rely on large datasets.}
For social media identity cloning, such weak labels can be obtained from training a DL classifier.

Let $\mathcal{D}=\left\{x_i,y_i\right\}^n_{i=1}$ denote a set of $n$ account pairs, with $\mathcal{X}=\left\{x_i\right\}^n_{i=1}$ denoting the accounts pair representation and $\mathcal{Y}=\left\{y_i\right\}^n_{i=1}\subset \left\{0,1\right\}^n$ denoting the corresponding clean labels indicating whether or not the account pair contains a cloned account and its victim. 
For the weak labels,
we predict the weak labels using the DF model described in the previous section (Sec \ref{df}). The weak labels are denoted as $\mathcal{\hat{D}}=\left\{\hat{x_j},\hat{y_j}\right\}^N_{j=1}$ where $\mathcal{\hat{X}}=\left\{\hat{x_j}\right\}^N_{j=1}$ denotes the $N$ unlabeled accounts pair representation and $\mathcal{\hat{Y}}=\left\{\hat{y_j}\right\}^N_{j=1}$ is the set of weak labels predicted from the DF model. 
\\(3) \textbf{Model Training}\\
Given a set of samples $\mathcal{D}=\left\{x_i,y_i\right\}^n_{i=1}$ with ground truth (clean) labels and a set of weak labels  $\mathcal{\hat{D}}=\left\{\hat{x_j},\hat{y_j}\right\}^N_{j=1}$, the DF model described in the previous section aims to predict a sample $x$. The label of a sample is predicted in Equation \ref{concat}.

\subsection{Cryptography Based Authentication}\label{cba}
We present a cryptography-based authentication protocol to verify if an account pair detected by the similar identify detection approach is created by the same user. 
Our protocol is built on the assumption that a newer account in a pair of similar accounts is more likely a cloned account according to the existing study. This assumption is supported by research showing that attackers usually choose well-established accounts as those accounts have more credibility and trust within the social network\cite{fire2014online}. 
Within this protocol, an authentication agent (AA) encrypts two random messages respectively using the older and newer accounts' identities and requests the newer account to successfully decrypt these encrypted messages for authentication. In the remaining of the subsection, we first introduce the preliminaries and describe the system model; next, we explain the proposed framework architecture and the adopted algorithms; finally, we analyze the convenient attacks during the authentication process and analyze how these attacks can be resolved with our solution. 

\subsubsection{Preliminaries}
This section provides a brief background introduction about Bilinear Maps, Identity-Based Encryption (IBE) and Advanced Encryption Standard (AES), which are the primary techniques from which the proposed solution is derived.
\\ (1) \textbf{Bilinear Maps} \\
Let $\mathbb{G}_1$ and $\mathbb{G}_2$ be two cyclic groups with a prime order $p$. A bilinear map $e: \mathbb{G}_1 \times \mathbb{G}_1 \rightarrow \mathbb{G}_2$ is a map with the following properties:
    
    \textbf{Bilinearity}: $e(g^{a}_{1},g^{b}_{2}) = e(g_{1},g_{2})^{ab}$ for all $a,b \in \mathbb{Z}_p$.
    
    \textbf{Non-degeneracy:} there exist $g_1, g_2 \in \mathbb{G}_1$ such that $e(g_{1},g_{2}) \neq 1$
    
    \textbf{Computability:} for all $a,b \in \mathbb{Z}_p$, an efficient algorithm exists to compute $e(g_{1},g_{2})$
\\(2) \textbf{Identity-Based Encryption (IBE)} \\
We adopt the Boneh-Franklin IBE scheme \cite{boneh2001identity} in our proposed solution. The Boneh-Franklin IBE scheme is provably secure and efficient \cite{gorantla2005survey}. The scheme consists of the following algorithms:
    
    \textbf{Setup: }The input of the setup algorithm is a security parameter $t$ and it outputs a public parameter ($PK$) and a master key ($MK$).
    
    \textbf{Extract: }The inputs of the extract algorithm include the public parameter ($PK$), the master key ($MK$) and an identity $(ID)$. It outputs a private key $d_{ID}$ for the identity.
    
    \textbf{Encrypt:} The encryption algorithm takes a randomly created message ($M$), the public parameter ($PK$) and the identity $(ID)$ as inputs. It outputs the encrypted message ($C$).
    
    \textbf{Decrypt:} The decryption algorithm takes the ciphertext ($C$), the public parameter ($PK$) and an identity $(ID)$ as inputs. It outputs the message ($M$).
\\(3) \textbf{Advanced Encryption Standard (AES)} \\
The AES algorithm is a well-known approach for symmetric key encryption. It performs encryption and decryption using a single key. AES uses a 128-bit block size and a 128-bit, 192-bit, or 256-bit key size. The key size for an AES cipher is determined by the number of AES rounds, which is 10, 12, or 14 for 128, 192, or 256-bit keys, respectively. The AES encryption process is operated on a 4×4 byte matrix that is called a state array. Each AES round contains several encryption processing steps \cite{aes_2001} which are described below: 
    
    \textbf{SubBytes: }A non-linear substitution phase in which each byte is substituted by another using a lookup table.
    
    \textbf{ShiftRows:} Each row of the 4x4 matrix is shifted cyclically by a predefined offset.
    
    \textbf{MixColumns:} A linear mixing operation that mixes the columns.
    
    \textbf{AddRoundKey: } XOR operation is performed between each byte of the state and the subkey, which is obtained from the main key using Rijndael's key schedule \cite{daemen1999aes} that converts a short key into a series of independent round keys. 

\subsubsection{System Design}
\begin{figure}[]
  \centering
  \includegraphics[width=0.6\columnwidth]{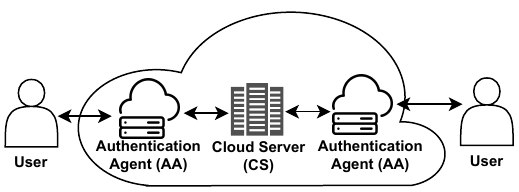}
  \caption{{System model}}
  \label{system_model}
\end{figure}

{Our proposed framework is an extension of the Boneh-Franklin IBE scheme \cite{boneh2001identity} that assumes that the private key generator (PKG) (i.e. AA) is a fully trusted entity. The major difference between the Boneh-Franklin IBE scheme and our proposed framework is that the former has only one AA and the latter has multiple AAs. Besides, both the AA and CS reside in the cloud side and are fully managed by the social-sensor cloud provider, which can be assumed as fully trusted \cite{wang2010enabling}. Our proposed framework for cryptography-based authentication is shown in Figure \ref{system_model}, including:}
      
      \textbf{Authentication Agent (AA):} Each AA is a fully trusted entity that is responsible for generating a private key for a new account. It is also responsible for delivering the private key to the new account through secure channels and encrypting a random message and decrypting an account's response for account authentication.
      
      \textbf{Cloud Server (CS):} The CS is also a fully trusted entity. It offers a cloud-based data storage solution. Social-sensor cloud providers deploy CS to facilitate the storage of  detected similar account pairs and their  authentication status.
    
    \textbf{Users:} The users obtain their private keys via registration. They are also required to authenticate themselves when there is are existing similar accounts.


\subsubsection{The Authentication Framework Architecture}\label{arch}
\begin{figure}[]
  \centering
  \includegraphics[width=0.8\columnwidth]{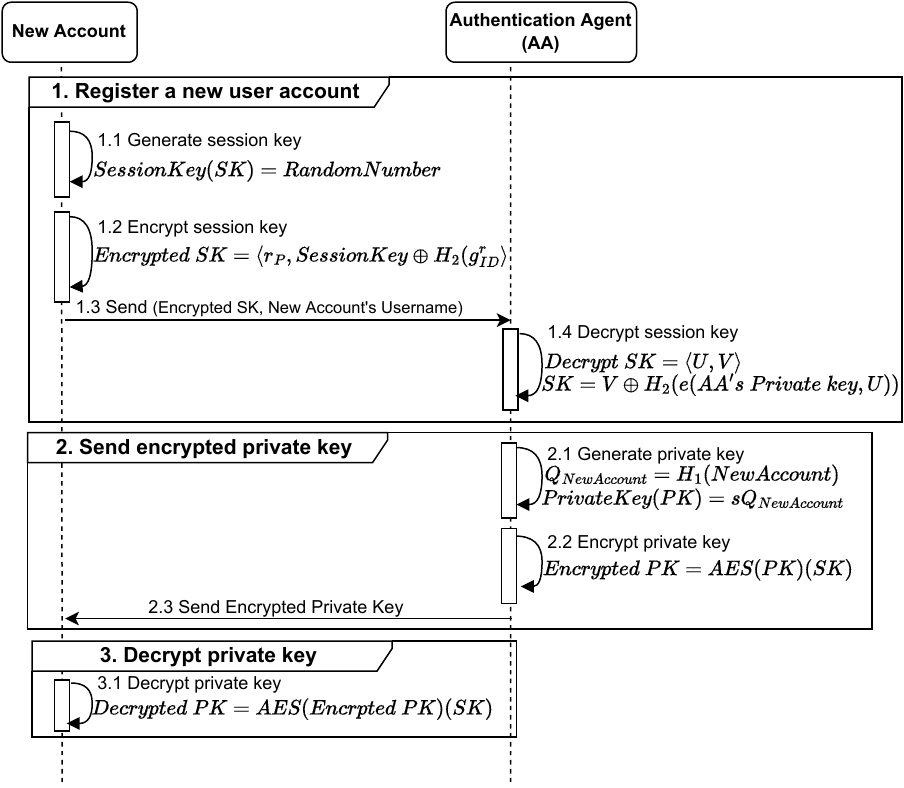}
  \caption{{A sequence diagram of the key generation process.}}
  \label{uml1}
\end{figure}

The proposed authentication protocol consists of two main parts: 1) key generation process; 2) authentication process.  
\paragraph{\textbf{Key Generation Process}}\label{kg}
Figure \ref{uml1} shows the sequence diagram of the key generation process. 
      
 \textbf{Register a new user account:} Every time when a new account is registered, the system will send the account information (i.e. the username) to an AA to receive a private key. During this process, this new account generates a session key. The session key is encrypted using the AA's public key (i.e. the AA's ID) with the Encrypt algorithm  defined in Section \ref{Sys_def}. Finally, the encrypted session key and the user's identity are sent to the AA to request a private key.

\textbf{Send encrypted private key:} Once the AA receives a private key request containing an encrypted session key and a username, it decrypts the session key by running the Decrypt algorithm depicted in Section \ref{Sys_def}. The AA then generates a private key for this registered account by running the Extract algorithm in Section \ref{Sys_def}. Next, the AA encrypts the private key using the session key by applying the AES-Enc$_{PK}$ algorithm in Section \ref{Sys_def}. Finally, the AA sends the encrypted private key to the registered account.

\textbf{Decrypt private key:} Once the registered account receives the encrypted private key, the registered account decrypts the private key by running the AES-Dec$_{PK}$ algorithm described in Section \ref{Sys_def}.
\begin{figure}[]
  \centering
  \includegraphics[width=1\columnwidth]{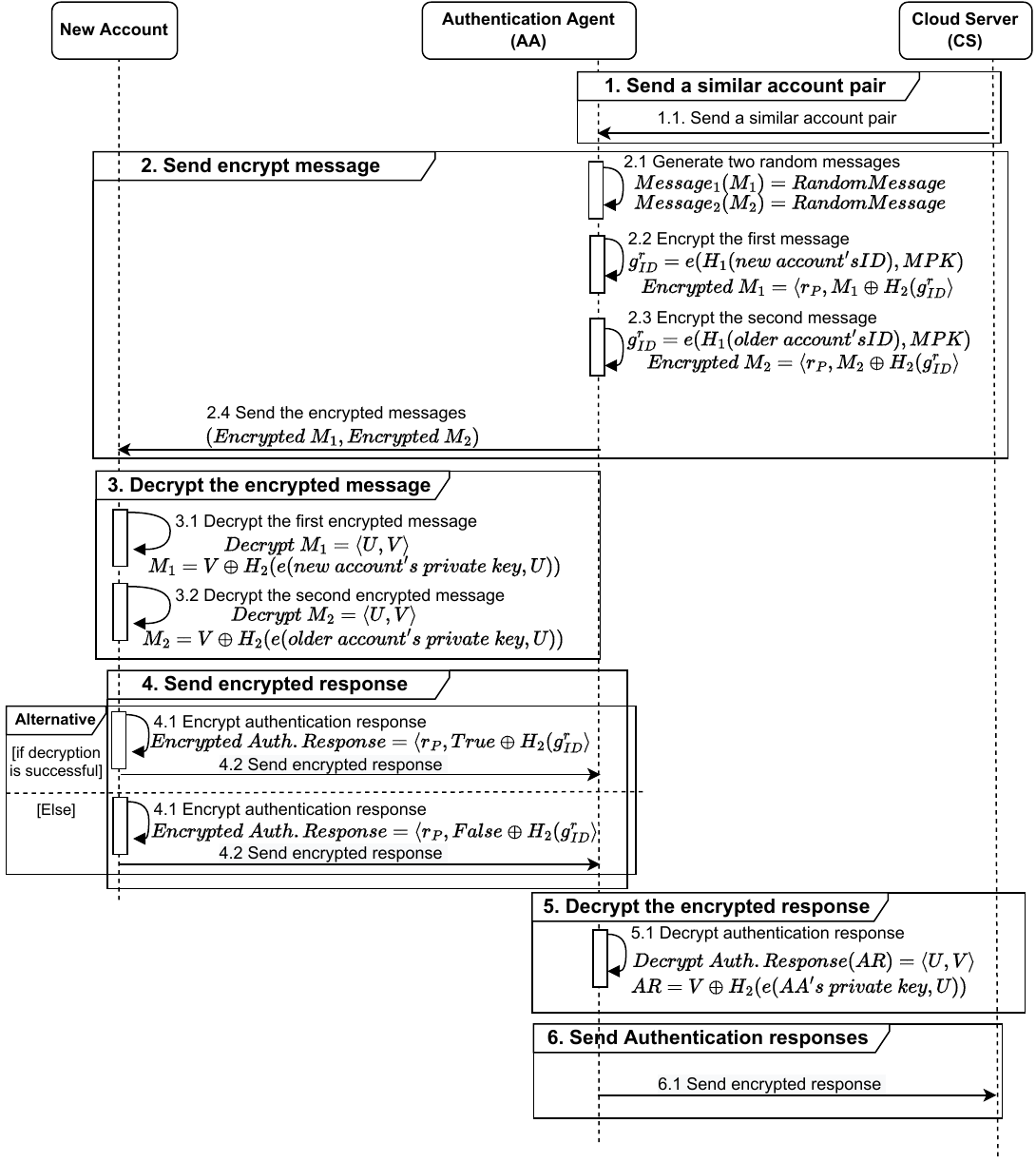}
  \caption{{A sequence diagram of the authentication process.}}
  \label{uml}
\end{figure}

\paragraph{\textbf{Authentication Process}}\label{a_p}
Figure \ref{uml} shows the sequence diagram of the authentication process. 

\textbf{Send a similar account pair:} Each time the CS detects a similar account pair whose authentication status is null, it will send the account pair information to an AA. The information includes the username and registration date of each account in the account pair. 

 \textbf{Send encrypted message:} Once the AA receives the information of a similar account pair, 
it checks the registration date of the account pair. {The AA then sends two encrypted messages to the newer account to request it to decrypt the messages for the authentication purpose.  The encrypted messages are two random messages respectively encrypted using the older and newer accounts' public keys (i.e. their usernames). The AA will run the Encrypt algorithm to encrypt the random messages. The encrypted messages can only be decrypted using their corresponding private keys}.

\textbf{Decrypt the encrypted message:} {Once the newer account receives the encrypted messages, the account user is required to decrypt the encrypted messages respectively with the older and newer accounts' private keys using the Decrypt algorithm. Here the responses can be categorized into two basic types: 1) The decryption is successful (i.e. both the messages are successfully decrypted), which indicates that the owner of the newer account owns the older account's private key and the account pair is created by the same user. 2) The decryption is unsuccessful (i.e. none or either of the messages are successfully decrpyted), which indicates that the newer account is possibly the clone of the older account or an attacker who steals the private key of the newer or older account.}

\textbf{Send encrypted response:} 
The response of the newer account is encrypted with the AA's ID using the Encrypt algorithm and 
sent to the AA.

 \textbf{Decrypt the encrypted response:} Once the AA receives the encrypted response from  the newer account, the AA decrypts this response with its private key by running the Decrypt algorithm.

 \textbf{Send authentication response:} The AA will send the decrypted response to the CS to update their authentication status (i.e. successful or unsuccessful).

Eventually, all the unsuccessfully authenticated account pairs are suspected to highly likely contain a cloned account and need further investigation.

\subsubsection{System Definition}\label{Sys_def}
We explain the algorithms adopted by the framework.

\textbf{Setup:} \label{setup_1} This algorithm only runs on the side of the AA. The AA selects a bilinear pairing $e:G \times G \rightarrow H$, where $G$ and $H$  are two cyclic groups of prime order $p$, with the size of $q$, where $q$ is prime power. The AA then selects a random value $s$, such that $s \in Z{^*_{q}}$\, and calculates $P_{pub} = sP$, where $P$ and $sP$ are point on elliptic curve.
Next, the AA employs the cryptography hash function $H_{1}:\left\{ 0,1\right\}^{*} \rightarrow G $, which allows the AA to map an identity $ID$ to an elliptic curve point. The AA also needs a hash function $H_{2}$ such that $H_{2}: H \rightarrow \left\{ 0,1\right\}^{n}$ to encrypt a message with size $n$. Finally, the system master public key (MPK) is $MPK = \langle e,n,G,H,P,P_{pub},H_{1},H_{2} \rangle$ and the system master private key (MSK) is $MSK = s \in Z{^*_{q}}$ for the AA. The AA uses the $MSK$ to generate a private key based on a user's $username$.

\textbf{Extract:} \label{extract_2} This algorithm also runs at the side of the AA only. A private key request $(Req_{PK})$ containing a user's $(username)$ and an encrypted session key using the AA's ID is sent to the AA. Once receiving the $Req_{PK}$, the AA maps the user's $username$  to a point $P$ on the elliptic curve $E$ as follows: $Q_{ID} = H_{1}(ID) \in G$, where $Q_{ID}$ is the result of mapping the user's $username$  to a point $P$. The user's private key $d_{ID}$ is set to $d_{ID} = sQ_{ID}$, where $s$ is the master private key of the AA.

 \textbf{AES-Enc$_{PK}$:} \label{AES-Enc_3} After the AA generates a private key for a given user, the AA uses Advanced Encryption Standard (AES) \cite{aes_2001} to encrypt the user's private key using the shared session key. The AA then sends the encrypted private key to this user.

 \textbf{AES-Dec$_{PK}$:} \label{AES-Dec_4} After the user receives his/her private key from the AA, the user uses AES to decrypt his/her private key using the shared session key.

\textbf{Encrypt:} \label{encrypt_5} 
The encryption algorithm selects a random integer $r \in Z{^*_{q}}$. It then uses the recipient's $username$ to calculate $Q_{ID} = H_{1}(ID) \in G$. Finally, it computes a ciphertext $C = \langle r_{P}, M \oplus H_{2}(g{^r_{ID}} \rangle$, where $g_{ID} = e(Q_{ID},MPK) \in H$. 

\textbf{Decrypt:} \label{decrypt_6} On receiving the ciphertext $C$ ($C = \langle U,V \rangle $), where $U$ and $V$ are the components of ciphertext $C$, 
the recipient  uses his/her private key $d_{ID}$ to decrypt $C$. The decrypted message $M$ is calculated  by $M = V \oplus H_{2} (e(d_{ID},U))$. 

\subsubsection{Attack Analysis}
\begin{figure}[]
  \centering
  \includegraphics[width=0.5\columnwidth]{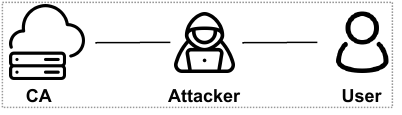}
  \caption{Secure channel attack model}
  \label{attack_model_a}
\end{figure}

We demonstrate how the proposed framework is proven to be secure and robust against potential attacks.
\paragraph{Channel Attacks}\label{s_r}
The proposed  protocol requires the AA to interact with a user account during
the key generation and authentication process. In such cases, an attacker can steal the information transmitted between the AA and the user account shown in Figure \ref{attack_model_a}. 
\paragraph{Prevention of Channel Attacks}
We prevent such an attack by encrypting the data transmitted between the user account and the AA in the following interactions:

\textbf{Private key generation:} 
With the proposed protocol,  each account first generates a session key that is only shared between the account and the AA. The session key is encrypted using the AA's ID (Step 1 in Section \ref{kg}). The AA then generates a private key for the account. Next, the AA encrypts the generated private key for the account using the shared session key (Step 2 in Section \ref{kg}). By encrypting the private key, this protocol provides a secure channel between the AA and the account for private key transmission. 

\textbf{Authentication:} {The random messages generated by the AA for authentication are respectively encrypted using the newer and older accounts' usernames (Step 2 in Section \ref{a_p}). The newer account is required to perform more secure double-key authentication (i.e. decrypting both the messages respectively using the newer and older accounts' private keys) to prove its identity. In addition,
 the authentication response of each account is encrypted using the AA's ID (Step 4 in Section \ref{a_p}). By encrypting the authentication messages and response, this protocol provides a secure channel for transmitting the authentication data between the AA and the account. The formal proof of the encryption and decryption was provided in \cite{boneh2001identity}}.

\section{Evaluation}\label{eval}
We conducted a set of experiments to verify the effectiveness of the proposed solution. The experiments will answer the following four key questions:
    
    \textbf{RQ1:} What are the impact of the components and the optimal hyperparameter values of the similar identity detection approach?
    
    \textbf{RQ2:} Does our solution outperform the state-of-the-art cloned identity detection approaches and other machine learning and deep learning-based models?
    
    \textbf{RQ3:} What is the required time efficiency of the cryptography-based authentication protocol?
    
    \textbf{RQ4:} What is the required key storage capacity of the cryptography-based authentication protocol?
\subsection{Experimental Environment}
All the experiments were conducted on a Ubuntu Server 20.04 LTS with Intel Core i5 1.80 GHz CPU and 16 GB RAM.  Furthermore, all the candidate models were implemented in Python. We extracted the pre-trained language representations utilised in similar identity detection using the SBERT package\footnote{https://github.com/UKPLab/sentence-transformers}. Additionally, we extracted the Node2Vec representations employed by  similar identity detection using the StellarGraph package\footnote{https://github.com/stellargraph/stellargraph}. We implemented the DL models evaluated using the Python-based Tensorflow\footnote{https://www.tensorflow.org/} library and the other machine learning models evaluated using scikit-learn\footnote{https://scikit-learn.org/stable/}. For the cryptography-based authentication, we used the Charm\footnote{https://github.com/JHUISI/charm} framework to implement Boneh-Franklin Identity based encryption. We also used PyCryptodome\footnote{https://github.com/Legrandin/pycryptodome} to implement AES based encryption. Finally, we implemented a prototype of the  cryptography-based authentication protocol, which can be downloaded from\footnote{https://github.com/a7madtalal123/Social-Sensor-Cloud-Service-Provider-Identity-Cloning-Detection}.
In addition, we executed all the machine and DL models 10 times with different random data combinations. The results are presented as a mean of all experiments conducted during the tests.

\subsection{Dataset}

To the best of our knowledge, Twitter is the only mainstream social media platform that has disclosed a set of cloned accounts\footnote{https://impersonation.mpi-sws.org/}. It contains 7,015 possibly cloned accounts and their targets (i.e. another set of 7,015 accounts). The majority of current research, albeit restricted in scope, evaluated proposed methodologies using simulated data. As a result, based on those Twitter accounts, we created a dataset to assess our proposed solution. We retrieved the non-privacy-sensitive user profile features of those accounts through Twitter APIs\footnote{https://developer.twitter.com/en/docs}. Further, most social media platforms claim that fake accounts (including cloned accounts) are a minority. For example, Twitter claims that fake accounts only account for about 5\% of the total accounts. Therefore, to mimic a real-world social media environment where cloned profiles are a minority, we randomly collected 500,000 public Twitter user accounts. Eventually, we developed a dataset that comprises a total of 514,030 public Twitter profiles\footnote{https://www.reuters.com/technology/twitter-estimates-spam-fake-accounts-represent-less-than-5-users-filing-2022-05-02/}.


\subsection{Other Methods Evaluated}
We evaluated and compared our proposed similar identity detection approach against a comprehensive set of existing identity cloning detection approaches listed below
\\ \textbf{Basic Profile Similarity (BPS) \cite{jin2011towards}:} This approach examines how much a specific user account and its presumed cloned account overlap using public features and similar friends.
\\ \textbf{Devmane and Rana \cite{devmane2014detection}:} This approach extracts user accounts' names, workplace, images, location, birthday, education, gender, and friends counts. It then compares these extracted features against a set of user accounts.
\\ \textbf{Goga et al. \cite{goga2015doppelganger}:} This technique extracts different user account features. It extracts user account public features, overlap account friends, overlap of account tweeting and differences between accounts. A linear kernel is then used to train an SVM classifier, which is subsequently used to identify whether a given account has been impersonated.
\\ \textbf{Kamhoua et al. \cite{kamhoua2017preventing}:} This technique evaluates the similarity of friend lists and calculates the similarity of features operating an adjusted similarity measure called Fuzzy-Sim. It examines the following features: name, city, friend list, place of employment age, gender and education. We utilised the same Fuzzy-Sim threshold values (i.e. 0.565 and 0.575) as indicated in the original paper.
\\ \textbf{NPS-AntiClone \cite{alharbi1234}:} This approach uses only the multi-view account representation of each account in the GC. It then determines the cosine similarity of the two accounts. Then, if the resemblance is more than 0.1, the account pair comprises the cloned account and its related target account.
\\ \textbf{Zheng et al. \cite{zheng2015detecting}:} This model is generally used to detect spammers. It utilises 18 different types of features such as profile-related (e.g. follower count)  and content-related (e.g. the average count of hashtags). Next, an SVM classifier based on a Radial Basis Function (RBF) kernel is trained to identify if a user account is a spammer user account.
\\ \textbf{DF$_{Clean}$ \cite{alharbi2021privacy}:} This approach is similar to our proposed similar identity detection approach. However, it only uses  clean labels to train the DF model.

Furthermore, we evaluated the proposed DF model against the listed below machine learning and DL models to validate the usage of the DF model as the classifier of the cloned identity. The current literature extensively employed the following models in the area of cloned identity detection in social media \cite{alharbi2021social}. These models include LR, RF, Adaboost (ADA), Deep Neural Network (DNN), K nearest neighbours (KNN), Support Vector Machine (SVM), Convolutional Neural Network (CNN) and Multi-layer Perceptron (MLP). Additionally, we evaluated the proposed DF model to variants of its base learners  (ERT-based DF (DF$_{ERT}$),  LR-based DF (DF$_{LR}$) and DF\(_{2RF,2ERT}\)) to demonstrate the model structure tuning process.

\subsection{Hyperparameter Tuning}
We fine-tuned all the hyperparameters of the supervised machine learning and DL models. Table \ref{parm} details the values of the hyperparameter utilised to configure the machine learning and DL models. Additionally, all the models are based on the non-privacy-sensitive user features obtained through Twitter APIs.

We configured the hyperparameters of the similar identity detection solution. The GC's optimal $\delta$ value is 0.8 according to the experimental results (see Section \ref{gc_v}). We utilised  `all-MiniLM-L6-v2'\footnote{https://huggingface.co/sentence-transformers/all-MiniLM-L6-v2} as the pre-trained model for SBERT. This pre-trained model was trained on millions of paraphrased sentences. SBERT's default dimension for its post representation is $385$. For the Node2Vec, we used its default dimension size ($128$ dimensions)  for the network view (i.e. follower and friend network). Additionally, we utilised $q = 2$ 
as the likelihood of moving away from the source node, $p = 0.5$ as the likelihood  of returning to the source node and $15$ as the maximum length of a random walk. We normalised all profile attribute views to $[0, 1]$. We set the wGCCA's weights $w$ to $[0.25, 0.5, 0.5, 0.25]$ which were determined in the experiment (see Section \ref{wgcca_w}).
\begin{table}[t]
\centering
\caption{Values of hyperparameter for ML/DL models}
\label{parm}
\resizebox{0.9\columnwidth}{!}{%
\begin{tabular}{p{0.1\columnwidth}p{0.7\columnwidth}}
\toprule
\textbf{Model} & \textbf{Parameter} \\ \toprule
ADA & estimators = 100 \\
RF & estimators = 50 \\
MLP & activation = relu, solver = adam \\
CNN & 8 layers, filters = 64 and 8, kernel size = 2 and 1, pool size = 2 \\
DNN & 6 layers (300, 250,150,100, 50, 1)\\
SVM & kernel = linear \\ 
KNN & neighbors = 15 \\ 
\toprule
\end{tabular}%
}
\end{table}

\subsection{Results and Discussion}
We used Precision, Recall and F1-Score in our evaluation. Precision is the ratio of accurately predicted account pairs (i.e. a cloned account and its related target), while Recall is the ratio of true account pairs that are accurately detected. 
\subsubsection{Impact of the components and hyperparameter (RQ1)}
\paragraph*{\textbf{Impact of the major components}} We evaluated the contribution of the key components of our proposed approach by comparing its performance with variants that excluded either the multi-view account representation (wGCCA), account pair feature representation, or weakly supervised labelling. The results demonstrate that integrating all three components significantly enhances detection accuracy, as they complement each other by capturing different aspects of the data. Table \ref{AbSt} presents the detailed comparison results.
\paragraph*{\textbf{Impact of the views}} To assess the contribution of the individual views in the multi-view account representation, we compared the performance of wGCCA-based detection (without account pair feature representation or weakly supervised labelling) using each view in isolation, as well as their combination. The analysis shows that combining all views—post view, network view, and profile attribute view—substantially improves the model's performance, highlighting the importance of a comprehensive multi-view approach. Table \ref{multiV} summarizes the comparison results.

\begin{table}[t]
\centering
\caption{Impact of the major components}
\label{AbSt}
\resizebox{\columnwidth}{!}{%
\begin{tabular}{p{0.65\columnwidth}p{0.2\columnwidth}}
\toprule
\textbf{Model} & \textbf{F1-Score(\%)} \\ \toprule
Our solution (w/o wGCCA) & 80.13 \\
Our solution (w/o Account pair feature representation)
& 78.32 \\
Our solution (w/o Weakly supervised labelling)
& 79.77 \\
Our solution (Combined) & \textbf{82.07} \\

\toprule
\end{tabular}%
}
\end{table}
\begin{table}[t]
\centering
\caption{Impact of the views on wGCCA based detection
}
\label{multiV}
\resizebox{0.9\columnwidth}{!}{%
\begin{tabular}{p{0.6\columnwidth}p{0.2\columnwidth}}
\toprule
\textbf{Model} & \textbf{F1-Score(\%)} \\ \toprule
wGCCA (Profile attributes) & 53.17 \\
wGCCA (Posts) & 55.17 \\
wGCCA (Network) & 61.81\\
Our wGCCA (All views)
& \textbf{71.89} \\  
\toprule
\end{tabular}%
}
\end{table}

\begin{figure}[t]
    \centering
\resizebox{0.6\columnwidth}{!}{%
\begin{tikzpicture} [thick,scale=1, every node/.style={scale=1}]
\begin{axis}[
    legend style={nodes={scale=0.6, transform shape}},
    width=8.5cm,
    height=5cm,
    title={},
    xlabel={$\delta$},
    ylabel={Performance (\%)},
    xmin=0.1, xmax=0.9,
    ymin=0, ymax=100,
    xtick={0.1,0.2,0.3,0.4,0.5,0.6,0.7,0.8,0.9},
    ytick={0,10,20,30,40,50,60,70,80,90,100},
    legend pos=south west,
    ymajorgrids=true,
    grid style=dotted,
]
\addplot[
    color=red,
    mark=*,
    mark size=2pt
    ]
    coordinates {
    (0.1,0.2424)(0.2,0.2416)(0.3,0.2394)(0.4,0.2354)(0.5,0.2319)(0.6,0.22917)(0.7,0.2568140)(0.8,0.302)(0.9,1.6582)
    
    };
    \addplot[
    color=black,
    mark=diamond*,
    mark size=2pt
    ]
    coordinates {
    (0.1,98.81)(0.2,98.50)(0.3,97.57)(0.4,95.95)(0.5,94.50)(0.6,93.12)(0.7,90.92)(0.8,88.12)(0.9,82.67)
    };
    \addplot[
    color=blue,
    mark=square*,
    mark size=2pt
    ]
    coordinates {
    (0.1,0.483723)(0.2,0.48219)(0.3,0.47765)(0.4,0.46973)(0.5,0.46269)(0.6,0.4572)(0.7,0.51218)(0.8,0.6027)(0.9,3.25130)
    };
\legend{Precision, Recall,F1-Score}
\end{axis}
\end{tikzpicture}}
    \caption{Impact of the threshold of GC}
    \label{impact_gc}
    \end{figure}
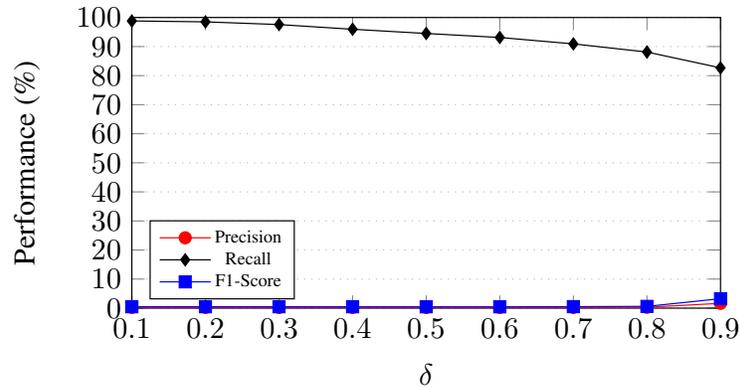

\begin{figure}[]
\centering
  
  \begin{tikzpicture}[thick,scale=1, every node/.style={scale=1}]
    	\definecolor{darkgreen}{rgb}{0.0, 0.2, 0.13}
    \begin{axis}[
        width=12cm,
        height=6cm,
        xtick={0,1,2,3,4,5,6,7,8,9},
        xticklabels={$_{[0.5,0.5,1,0.25]}$,$_{[1.0,1.0,1.0,1.0]}$,$_{[0.25,0.5,0.5,0.25]}$,$_{[0.5,0.5,0.5,0.5]}$,$_{[0.25,0.5,0.25,0.25]}$,$_{[1.0,1.0,0.5,0.25]}$,$_{[0.25,1.0,0.5,0.25]}$,$_{[0.25,0.25,0.25,0.25]}$,$_{[0.25,0.5,0.25,0.5]}$,$_{[0.25,1.0,1.0,0.25]}$},
        x tick label style={rotate=40,anchor=east},
        ytick={45,50,55,60,65,70,75,80,85},
        ymin=60,
        ymax=85,
        ymajorgrids=true,
        grid style=dotted,
        ylabel={F1-Score(\%)},
        every axis plot/.append style={
          ybar,
          bar width=8,
          bar shift=0pt,
          fill
        }
      ]
      \addplot [darkgreen] coordinates {(0,81.58)};
      \addplot [darkgreen] coordinates {(1,76.10)};
      \addplot [darkgreen] coordinates {(2,82.07)};
      \addplot [darkgreen] coordinates {(3,75.14)};
      \addplot [darkgreen] coordinates {(4,71.84)};
      \addplot [darkgreen] coordinates {(5,79.22)};
      \addplot [darkgreen] coordinates {(6,76.61)};
      \addplot [darkgreen] coordinates {(7,75.05)};
      \addplot [darkgreen] coordinates {(8,71.81)};
      \addplot [darkgreen] coordinates {(9,81.57)};

    \end{axis}
  \end{tikzpicture}
    \caption{Impact of the wGCCA's weight $w$}
  \label{weight}
\end{figure}
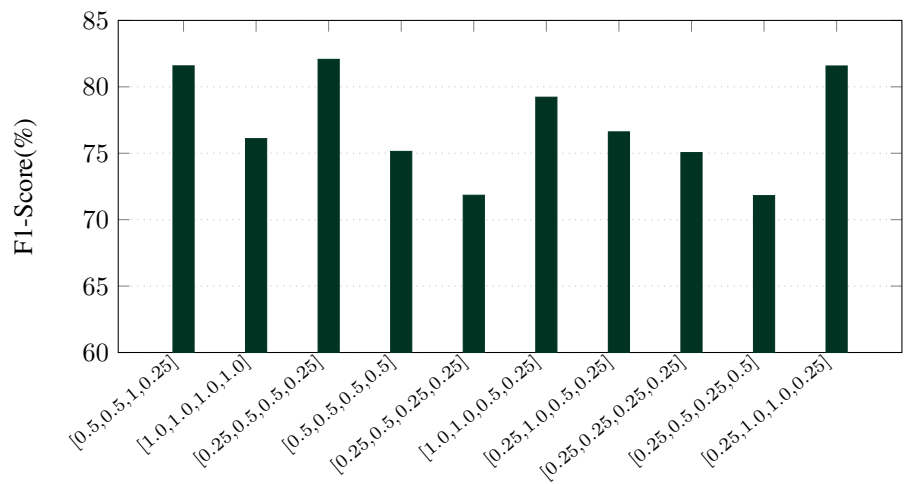

\paragraph*{\textbf{Impact of the threshold of GC}}\label{gc_v}
We analyzed the impact of the GC threshold $(\delta)$, as shown in Figure \ref{impact_gc}. It can be seen that Recall increases when the $\delta$ score was gradually declined from 0.8 to 0.1.  This means that, when the similarity score is lowered, the GC located more cloned account and victim account pairs. However, Precision is very low. We attribute that the GC located more irrelevant account pairs. This would remarkably impact the efficiency of the proposed solution.  Therefore, we choose 0.8 since it locates most possibly cloned account pairs with a relatively lower number of similar account pairs.
\paragraph*{\textbf{Impact of the wGCCA's weight $w$}}\label{wgcca_w}
We conducted a set of experiments to determine the effect of the wGCCA's weight $w$ on each view (i.e. post, network and profile views). To perform this, we examined several weight combinations (i.e. 0.25, 0.5, and 1) for each view. Each view was given a weight \textit{$[$post, friend, follower, attribute$]$}. The top 10 outcomes obtained against various weight combinations are shown in Figure \ref{weight}. We make the following observations. The F1-Score was increased when the profile attribute view was assigned with lower weights. Next, the post view did not impose much impact on the wGCCA embedding. In contrast, the network view (i.e. friends and followers) had a high impact. The F1-Score was risen when the weight of network view was increased. We finally chose $[0.25,0.5,0.5,0.25]$ as the optimal values of $w$ for each view.


\paragraph*{\textbf{Impact of the DF structure}}
We performed a series of experiments to analyze the impact of the DF structure. To this end, we tested various structures of the DF. Table \ref{imapct_DF} shows the performance results. Our proposed DF\(_{2RF, ERT, LR}\) (DF$_{Clean}$ \cite{alharbi2021privacy}) structure achieved 80.82\%, 78.76\% and 79.77\% on Precision, Recall and  F1-Score,  respectively. Employing a single type of classifier did not produce satisfactory performance results. Our proposed DF structure generates a cascading ensemble of these basic learners (i.e. ERT, LR and RF), in which each cascade consists of the aforesaid basic learners. In this regard, 
the cascading ensemble 
models the diversity of learning functions, which improves its generalisation performance and thus ultimately leads to a significant performance improvement.

\begin{figure}[t]
    \centering
\resizebox{0.6\columnwidth}{!}{%
\begin{tikzpicture} [thick,scale=1, every node/.style={scale=1}]
\begin{axis}[
    legend style={nodes={scale=0.6, transform shape}},
    width=8.5cm,
    height=4.5cm,
    title={},
    xlabel={Weak labels (\%)},
    ylabel={F1-Score (\%)},
    xmin=60, xmax=100,
    ymin=65, ymax=90,
    xtick={60,70,80,90,100},
    ytick={65,70,75,80,85,90},
    legend pos=south east,
    ymajorgrids=true,
    grid style=dotted,
]
\addplot[
    color=red,
    mark=*,
    mark size=2pt
    ]
    coordinates {
    (60,79.18334)(70,74.251211)(80,78.058543)(90,71.251761)(100,82.070378)
    
    };
\legend{F1-Score}
\end{axis}
\end{tikzpicture}}
    \caption{Impact of the weak labels percentage}
    \label{weak_l}
    \end{figure}
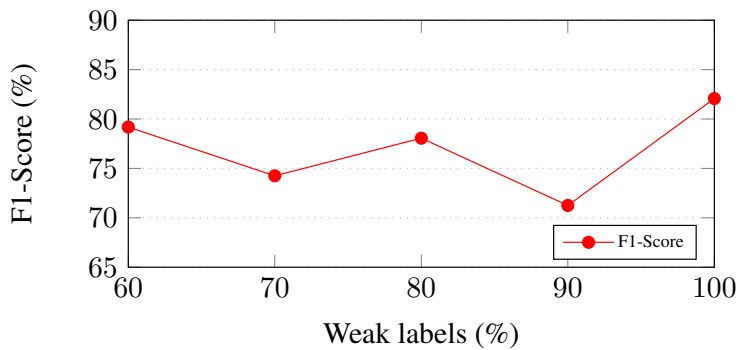

\begin{table}[t]
\centering
\caption{Impact of the DF structure
}
\label{imapct_DF}
\resizebox{0.9\columnwidth}{!}{%
\begin{tabular}{p{0.25\columnwidth}p{0.15\columnwidth}p{0.15\columnwidth}p{0.18\columnwidth}}
\toprule
\textbf{Model} &   \textbf{Precision(\%)} & \textbf{Recall(\%)} & \textbf{F1-Score(\%)}  \\ \toprule
DF\(_{RF}\)&  83.13&	73.35&	77.65  \\
DF\(_{ERT}\)&  49.73&	68.98&	56.09  \\
DF\(_{LR}\)&  83.83&	30.32&	44.49  \\
DF\(_{2RF,2ERT}\) &  \textbf{84.26}&	72.49&	77.89  \\
DF\(_{2RF, ERT, LR}\) (DF$_{Clean}$ \cite{alharbi2021privacy})   &   80.82 &	\textbf{78.76} &	\textbf{79.77}   \\ 
\toprule
\end{tabular}%
}
\end{table}
\paragraph*{\textbf{Impact of the weak label percentage}}
In real scenarios, the amount of ground truth data is usually limited.
On the other hand, a large amount of unlabeled data is available. In this experiment, we analyzed the impact of the weak labels percentage on our similar identity detection. To this end, we varied the weak label proportions between 60\% and 100\%. Figure \ref{weak_l} shows the performance results on those weak label percentages. It can be seen that the 100\% weak labels generate the highest F1-Score. This again validates the performance of the weak labels on improving the robustness of the trained model. Accordingly, we trained the DF model purely on all the weakly labelled data regenerated by the same DF model.

\subsubsection{Performance of the proposed similar identity detection approach (RQ2)} \label{RQ2_result}
Table \ref{per} shows the result of the performance comparison. Our proposed approach outperformed all the existing state-of-the-art identity detection approaches in terms of Precision and F1-Score. 
In addition, our proposed DF model trained on  weak labels achieves better performance than the DF model only trained on clean labels on the same metrics. This shows the robustness of the weak label based training. However, the recall rate of the DF drops when being trained on weak labels. In this regard, the weakly labelled training data contain noises that  affect the performance of the model on retrieving the number of true cloned account pairs.
The techniques proposed by Kamhoua et al. \cite{kamhoua2017preventing} and Devmane and Rana \cite{devmane2014detection} performed poorly since they use simple similarity techniques to calculate the similarity of profile features. They also did not take into account the post and network views of  users. BSP \cite{jin2011towards} is also based on simple profile attribute similarity and shared friends between account pairs. Similarly, Goga et al. \cite{goga2015doppelganger} employed a conventional similarity method that is based on human-defined measures. 
Zheng et al. \cite{zheng2015detecting}'s method entirely focuses  on spammer detection. It detects whether or not an account is a spammer by analysing its spammer behaviour patterns. However, cloned profiles' behaviour is different from spammers. Cloned profiles attempt to replicate the behaviour of genuine profiles and  thus are difficult to be detected by spammer detection techniques. The results show that our proposed similar identity detection approach better fits the scenario in which only the non-privacy-sensitive user profile features are employed to detect identity cloning. 
\begin{table}[t]
\centering
\caption{Comparison with the existing state-of-the-art identity detection techniques.}
\label{per}
\resizebox{\columnwidth}{!}{%
\begin{tabular}{p{0.34\columnwidth}p{0.15\columnwidth}p{0.15\columnwidth}p{0.18\columnwidth}}
\toprule
\textbf{Method} &   \textbf{Precision(\%)} & \textbf{Recall(\%)} & \textbf{F1-Score(\%)}  \\ \toprule
BSP \cite{jin2011towards} & 68.31	&75.14&	71.56   \\
Devmane and Rana \cite{devmane2014detection}&  64.31 &	77.14 &	70.15   \\
Goga et al. \cite{goga2015doppelganger} & 65.85&	73.74&	69.57   \\ 
Kamhoua et al. \cite{kamhoua2017preventing}  & 60.17 &	76.66 &	67.42     \\
NPS-AntiClone \cite{alharbi1234} & 71.14 & 67.67 & 69.36 \\
Zheng et al. \cite{zheng2015detecting} &  70.75&	71.47&	71.11 \\
DF$_{Clean}$ \cite{alharbi2021privacy} & 80.82 & \textbf{78.76} & 79.77 \\
Our solution &    \textbf{91.04} &	74.71 &	\textbf{82.07}  \\ 
\toprule
\end{tabular}%
}
\end{table}

Table \ref{ml_per} shows the comparison results between the proposed DF model and the other machine learning and DL models. It can be seen that our DF model notably surpassed all of the other candidate models on all the three metrics.  
The DF can automatically adjust the model complexity to adapt to the actual problem context. With this feature, the generation of DF models tends to be considerably simpler to handle and the quality of the generated models appear to be more adaptable to the given problem  compared with the other models (e.g. DNNs) \cite{zhou2017deep}. 

\begin{table}[t]
\centering
\caption{Comparison with the candidate machine learning and DL models evaluated as the predictor of our proposed similar account detection technique 
}
\label{ml_per}
\resizebox{0.9\columnwidth}{!}{%
\begin{tabular}{p{0.25\columnwidth}p{0.15\columnwidth}p{0.15\columnwidth}p{0.18\columnwidth}}
\toprule
\textbf{Model} &   \textbf{Precision(\%)} & \textbf{Recall(\%)} & \textbf{F1-Score(\%)}  \\ \toprule
ADA   &  83.17&	41.42&	52.47  \\
CNN   &  89.47&	58.65&	70.48  \\
DNN   &  90.87&	43.94&	58.01  \\
KNN   &  89.40&	49.66&	63.61  \\
LR   &  86.74&	22.33&	34.34  \\
SVM &  85.39&	28.71&	42.93  \\
MLP &  81.51&	64.25&	71.84  \\
RF &  78.36&	64.51&	70.27  \\ 
Our solution  &   \textbf{91.04} &	\textbf{74.71} &	\textbf{82.07}   \\ 
\toprule
\end{tabular}%
}
\end{table}

\subsubsection{Time efficiency (RQ3)}
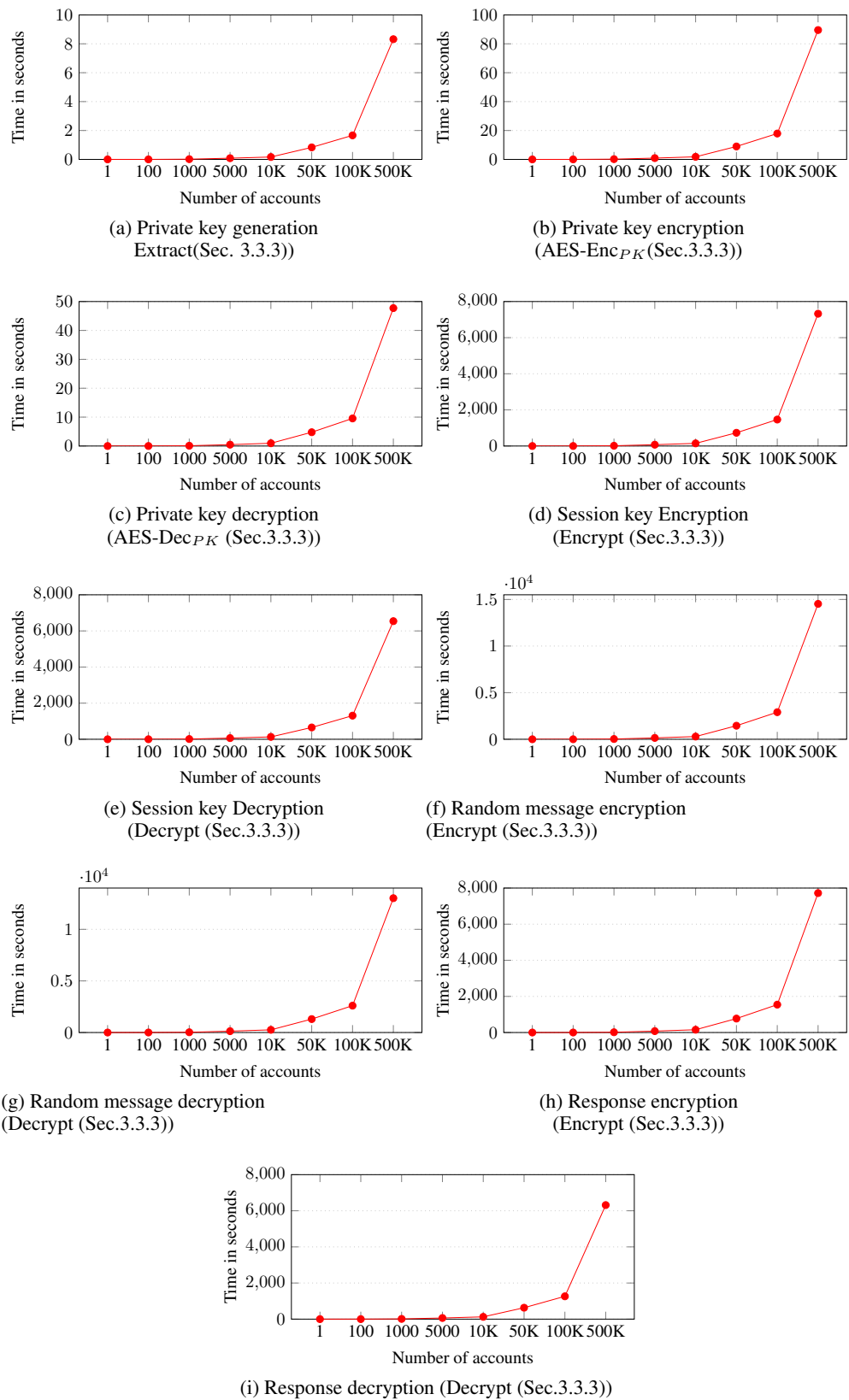
\begin{figure}[t]
    \centering
    \subfloat[][Private key generation \\ Extract(Sec. \ref{kg}))]{\resizebox{0.4\textwidth}{!}{
    \begin{tikzpicture} [thick,scale=1, every node/.style={scale=1}]
    \begin{axis}[
    width=8.5cm,
    height=4.5cm,
    symbolic x coords={0,1,100,1000,5000,10K,50K,100K,500K},
    ytick = {0,2,4, 6, 8,10},
    xtick=data,
    ymajorgrids=true,
    grid style=dotted,
    ymin=0, ymax=10,
    xlabel={Number of accounts},
    ylabel={Time in seconds},
    ]
    \addplot [color=red,
    mark=*,
    mark size=2pt] coordinates {
        (1,0.00113584310000001)(100,0.001664092)(1000,0.016640918)(5000,0.083204589)(10K,0.166409179)(50K,0.832045893)(100K,1.664091786)(500K,8.32045893)
    };
]
\end{axis}
\end{tikzpicture}
    }}
    \centering
        \subfloat[][Private key encryption \\ (AES-Enc$_{PK}$(Sec.\ref{kg}))]{\resizebox{0.4\textwidth}{!}{
    \begin{tikzpicture} [thick,scale=1, every node/.style={scale=1}]
    \begin{axis}[
    width=8.5cm,
    height=4.5cm,
    symbolic x coords={0,1,100,1000,5000,10K,50K,100K,500K},
    xtick=data,
    ymajorgrids=true,
    grid style=dotted,
    ytick = {0,20,40,60,80,100},
    ymin=0, ymax=100,
    xlabel={Number of accounts},
    ylabel={Time in seconds},
    ]
    \addplot [color=red,
    mark=*,
    mark size=2pt] coordinates {
        (1,0.000179050999999985)(100,0.0179051)(1000,0.179051)(5000,0.895255)(10K,1.79051)(50K,8.95255)(100K,17.9051)(500K,89.5255)
    };
]
\end{axis}
\end{tikzpicture}
    }}
    
    \centering
        \subfloat[][Private key decryption \\ (AES-Dec$_{PK}$ (Sec.\ref{kg}))]{\resizebox{0.4\textwidth}{!}{
    \begin{tikzpicture} [thick,scale=1, every node/.style={scale=1}]
    \begin{axis}[
    width=8.5cm,
    height=4.5cm,
    symbolic x coords={0,1,100,1000,5000,10K,50K,100K,500K},
    xtick=data,
    ymajorgrids=true,
    grid style=dotted,
    ytick = {0,10,20,30,40,50},
    ymin=0, ymax=50,
    xlabel={Number of accounts},
    ylabel={Time in seconds},
    ]
    \addplot [color=red,
    mark=*,
    mark size=2pt] coordinates {
        (1,0.0000954509999999686)(100,0.0095451)(1000,0.095451)(5000,0.477255)(10K,0.95451)(50K,4.77255)(100K,9.5451)(500K,47.7255)
    };
]
\end{axis}
\end{tikzpicture}
    }}
    \centering
        \subfloat[][Session key Encryption \\ (Encrypt (Sec.\ref{kg}))]{\resizebox{0.4\textwidth}{!}{
    \begin{tikzpicture} [thick,scale=1, every node/.style={scale=1}]
    \begin{axis}[
    width=8.5cm,
    height=4.5cm,
    symbolic x coords={0,1,100,1000,5000,10K,50K,100K,500K},
    xtick=data,
    ymajorgrids=true,
    grid style=dotted,
    ymin=0, ymax=8000,
    ytick = {0,2000,4000,6000,8000},
    xlabel={Number of accounts},
    ylabel={Time in seconds},
    ]
    \addplot [color=red,
    mark=*,
    mark size=2pt] coordinates {
        (1,0.014650719)(100,1.46507188)(1000,14.6507188)(5000,73.253594)(10K,146.507188)(50K,732.53594)(100K,1465.07188)(500K,7325.3594)
    };
]
\end{axis}
\end{tikzpicture}
    }}
    
        \centering
        \subfloat[][Session key Decryption \\ (Decrypt (Sec.\ref{kg}))]{\resizebox{0.4\textwidth}{!}{
    \begin{tikzpicture} [thick,scale=1, every node/.style={scale=1}]
    \begin{axis}[
    width=8.5cm,
    height=4.5cm,
    symbolic x coords={0,1,100,1000,5000,10K,50K,100K,500K},
    xtick=data,
    ymajorgrids=true,
    grid style=dotted,
    ymin=0, ymax=8000,
    xlabel={Number of accounts},
    ylabel={Time in seconds},
    ]
    \addplot [color=red,
    mark=*,
    mark size=2pt] coordinates {
        (1,0.013080489)(100,1.30804894)(1000,13.0804894)(5000,65.402447)(10K,130.804894)(50K,654.02447)(100K,1308.04894)(500K,6540.2447)
    };
]
\end{axis}
\end{tikzpicture}
    }}
            \centering
        \subfloat[][Random message encryption \\ (Encrypt (Sec.\ref{a_p}))]{\resizebox{0.4\textwidth}{!}{
    \begin{tikzpicture} [thick,scale=1, every node/.style={scale=1}]
    \begin{axis}[
    width=8.5cm,
    height=4.5cm,
    symbolic x coords={0,1,100,1000,5000,10K,50K,100K,500K},
    xtick=data,
    ymajorgrids=true,
    grid style=dotted,
    ymin=0, ymax=15500,
    xlabel={Number of accounts},
    ylabel={Time in seconds},
    legend style={at={(0.1,0.9)},anchor=north west},
    ]
    \addplot [color=red,
    mark=*,
    mark size=2pt] coordinates {
        (1,0.029050537)(100,2.90505372)(1000,29.0505372)(5000,145.252686)(10K,290.505372)(50K,1452.52686)(100K,2905.05372)(500K,14525.2686)
    };
        
]

\end{axis}
\end{tikzpicture}
    }}
                
                \centering
        \subfloat[][Random message decryption \\ (Decrypt (Sec.\ref{a_p}))]{\resizebox{0.4\textwidth}{!}{
    \begin{tikzpicture} [thick,scale=1, every node/.style={scale=1}]
    \begin{axis}[
    width=8.5cm,
    height=4.5cm,
    symbolic x coords={0,1,100,1000,5000,10K,50K,100K,500K},
    xtick=data,
    ymajorgrids=true,
    grid style=dotted,
    ymin=0, ymax=14000,
    xlabel={Number of accounts},
    ylabel={Time in seconds},
    legend style={at={(0.1,0.9)},anchor=north west},
    ]
    \addplot [color=red,
    mark=*,
    mark size=2pt] coordinates {
        (1,0.02603425)(100,2.603425)(1000,26.03425)(5000,130.17125)(10K,260.3425)(50K,1301.7125)(100K,2603.425)(500K,13017.125)
    };
]
\end{axis}
\end{tikzpicture}
    }}
                    \centering
        \subfloat[][Response encryption \\ (Encrypt (Sec.\ref{a_p}))]{\resizebox{0.4\textwidth}{!}{
    \begin{tikzpicture} [thick,scale=1, every node/.style={scale=1}]
    \begin{axis}[
    width=8.5cm,
    height=4.5cm,
    symbolic x coords={0,1,100,1000,5000,10K,50K,100K,500K},
    xtick=data,
    ymajorgrids=true,
    grid style=dotted,
    ymin=0, ymax=8000,
    xlabel={Number of accounts},
    ylabel={Time in seconds},
    ]
    \addplot [color=red,
    mark=*,
    mark size=2pt] coordinates {
        (1,0.015441534)(100,1.5441534)(1000,15.441534)(5000,77.20767)(10K,154.41534)(50K,772.0767)(100K,1544.1534)(500K,7720.767)
    };
]
\end{axis}
\end{tikzpicture}
    }}
    
            \subfloat[][Response decryption  (Decrypt (Sec.\ref{a_p}))]{\resizebox{0.4\textwidth}{!}{
    \begin{tikzpicture} [thick,scale=1, every node/.style={scale=1}]
    \begin{axis}[
    width=8.5cm,
    height=4.5cm,
    symbolic x coords={0,1,100,1000,5000,10K,50K,100K,500K},
    xtick=data,
    ymajorgrids=true,
    grid style=dotted,
    ymin=0, ymax=8000,
    xlabel={Number of accounts},
    ylabel={Time in seconds},
    ]
    \addplot [color=red,
    mark=*,
    mark size=2pt] coordinates {
        (1,0.012631537)(100,1.2631537)(1000,12.631537)(5000,63.157685)(10K,126.31537)(50K,631.57685)(100K,1263.1537)(500K,6315.7685)
    };
]
\end{axis}
\end{tikzpicture}
    }}
    \caption{Time efficiency}
    \label{Time_graph}
    \end{figure}

In this experiment, we analyzed the time consumption required by the cryptography-based authentication protocol. Our proposed authentication protocol contains a number of algorithms (see Section \ref{Sys_def}). Thus, we evaluated the time cost of each algorithm when being applying to the protocol (see Section \ref{arch}). The efficiency of our proposed authentication protocol is shown in Figure \ref{Time_graph}. Figure \ref{Time_graph}a shows the time consumption for extracting a private key versus the number of users. It approximately takes around 8s for 500k accounts. In Figure \ref{Time_graph}b, we can see the time (approximately 90s for 500k accounts) taken for encrypting the user's private key while in Figure \ref{Time_graph}c we can see the time (less than 50s for those accounts) consumed for decrypting a user's private key. Figure \ref{Time_graph}d shows the average encryption time of a session key of a user is around 0.01s. In Figure \ref{Time_graph}e, the average time taken for decrypting a session key of a user is  0.013s. {Figure \ref{Time_graph}f and g show the encryption and decryption time of random messages. It takes around 0.029s and 0.026s on average to respectively encrypt and decrypt two messages.}
In Figure \ref{Time_graph}h and \ref{Time_graph}i, we can see that the  encryption of an account's responses takes 0.015s approximately and its decryption  takes 0.012s roughly.
Overall, it can be seen that the time costs of all these algorithms are increased linearly along with the rising number of accounts. Thus, it means that the time consumption growth is always kept at a relatively fixed rate.  The evaluation results show that our proposed authentication protocol is computationally efficient.
    
\subsubsection{Key storage (RQ4)}
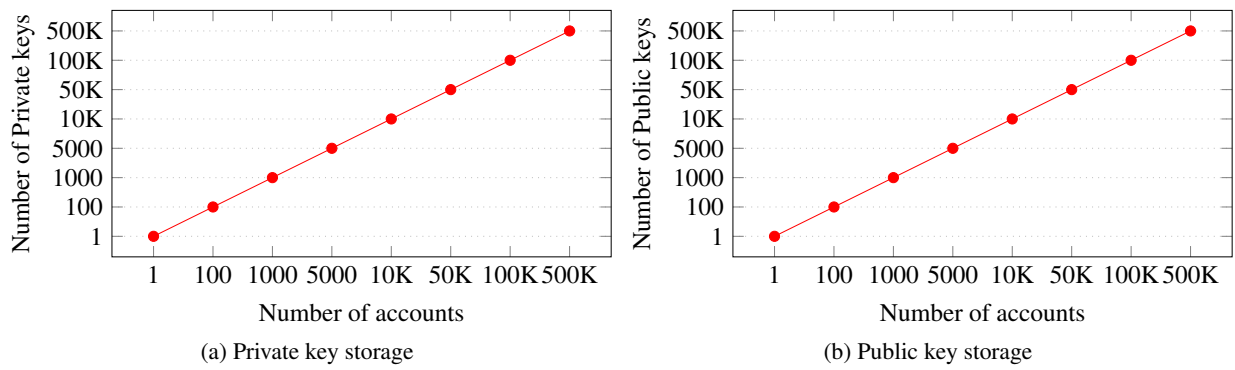
\begin{figure}[t]
    \centering
    \subfloat[][Private key storage]{\resizebox{0.5\columnwidth}{!}{
    \begin{tikzpicture} [thick,scale=1, every node/.style={scale=1}]
    \begin{axis}[
    width=8.5cm,
    height=5cm,
    symbolic x coords={0,1,100,1000,5000,10K,50K,100K,500K},
    symbolic y coords={0,1,100,1000,5000,10K,50K,100K,500K},
    ytick = {0,1,100,1000,5000,10K,50K,100K,500K},
    xtick=data,
    ymajorgrids=true,
    grid style=dotted,
    xlabel={Number of accounts},
    ylabel={Number of Private keys},
    ]
    \addplot [color=red,
    mark=*,
    mark size=2pt] coordinates {
        (1,1)(100,100)(1000,1000)(5000,5000)(10K,10K)(50K,50K)(100K,100K)(500K,500K)
    };
]
\end{axis}
\end{tikzpicture}
    }}
    \centering
        \subfloat[][Public key storage]{\resizebox{0.5\columnwidth}{!}{
    \begin{tikzpicture} [thick,scale=1, every node/.style={scale=1}]
    \begin{axis}[
    width=8.5cm,
    height=5cm,
    symbolic x coords={0,1,100,1000,5000,10K,50K,100K,500K},
    symbolic y coords={0,1,100,1000,5000,10K,50K,100K,500K},
    ytick = {0,1,100,1000,5000,10K,50K,100K,500K},
    xtick=data,
    ymajorgrids=true,
    grid style=dotted,
    xlabel={Number of accounts},
    ylabel={Number of Public keys},
    ]
    \addplot [color=red,
    mark=*,
    mark size=2pt] coordinates {
        (1,1)(100,100)(1000,1000)(5000,5000)(10K,10K)(50K,50K)(100K,100K)(500K,500K)
    };
]
\end{axis}
\end{tikzpicture}
    }}
    \caption{Key storage}
    \label{key_sto}
    \end{figure}

We analyzed the storage capacity required by storing the keys for the cryptography-based authentication protocol.  The storage of the private and public keys required by authentication in our dataset is shown in Figure \ref{key_sto}. 
 It can be seen from Figure \ref{key_sto}a that only one private key is required for the newly registered account at registration time. The length of each account's private key is 1024 bits. Similarly, a public key is required for each account and  is the account's username in this case, as shown in Figure \ref{key_sto}b. Overall, it can be observed that the required key storage capacity is increased linearly. The result of the required key storage capacity shows that our proposed authentication protocol provides very effective key storage solution.
    
\subsection{Summary of Findings}
We can make the following conclusions based on the aforementioned experimental analysis: 1) Our proposed similar identity detection approach outperformed the state-of-the-art-identity cloning detection methods. Additionally, it outperformed other machine and DL models (e.g. RF, DNN, etc). 2) Our proposed cryptography-based authentication is proved to be efficient in terms of time efficiency requirement and effective in terms of storage requirement.  


\section{Conclusion and Future Work}\label{concl}
We proposed a technique for detecting identity cloning in SocSen services with two components: 1) detecting similar accounts using non-sensitive profile data and a deep forest model, and 2) verifying if similar accounts belong to the same user by requiring the newer account to authenticate with the older account's private key.

Our future work will aim to explore additional datasets as they become available or develop methods to augment our existing data to further validate our approach. We also plan to integrate large language models (LLMs), which could further complement our proposed similar identity detection approach by enhancing the analysis of user-generated content, such as posts and descriptions. 
In addition, we will investigate how models like XGBoost and LightGBM could complement or enhance our approach, especially in terms of detection accuracy and computational efficiency. We plan to explore advanced techniques for refining weak label generation, such as incorporating confidence-based filtering and active learning strategies, to further enhance the robustness of our model against noisy labels and improve its overall performance.
Finally, we target making real-world case studies to evaluate the proposed cryptography-based authentication approach in practice.
\bibliographystyle{plain}
\bibliography{ref}
\end{document}